\documentclass[12pt]{article}
\usepackage{graphicx}
\usepackage{subfigure}

\newcommand{\beq}{\begin{equation}}
\newcommand{\eeq}{\end{equation}}
\newcommand{\bea}{\begin{eqnarray}}
\newcommand{\eea}{\end{eqnarray}}

\newcommand{\cM}{{\cal M} }

\newcommand{\CO}{{\cal O} }

\begin{document}
\topmargin 0pt
\oddsidemargin 5mm
\headheight 0pt
\topskip 0mm

\addtolength{\baselineskip}{0.20\baselineskip}

\pagestyle{empty}

\begin{center}

\vspace{18pt}
{\large \bf Spin Resolution of Glueballs in 2+1 Dimensional Lattice Gauge Theory}

\vspace{2 truecm}

{\sc Robert W. Johnson\footnote{e-mail: r.johnson1@physics.ox.ac.uk}}

\vspace{1 truecm}

{\em Department of Physics, University of Oxford \\
Theoretical Physics,\\
1 Keble Road,\\
 Oxford OX1 3NP, UK\\}

\vspace{3 truecm}

\end{center}

\noindent

{\bf Abstract.} Conventional lattice gauge theory assigns the lowest
spin compatible with the symmetry channel of a given operator to the
state coupling to that operator.  Operators on a cubic lattice, however,
are only defined on angles of $\pi/2$, hence states with spin equal
{\it{modulo}} 4 may overlap significantly.  This paper explores a new
technique for generating lattice operators that may be placed onto the
lattice at angles other than $\pi/2$, thereby resolving this
{\it{modulo}} 4 ambiguity.  Calculations of the mass of states with spin equal to 0, 2, and 4 are performed in the positive parity and charge conjugation channel and compared to the spectrum from previous lattice calculations.  These masses compare well for spin 0 and 2, and for spin 4 the mass agrees with a state conventionally assigned spin 0, raising the possibility of mis-identification of the spin of states coupling to some traditional operators.

\vfill
\begin{flushleft}
\end{flushleft}
\newpage
\setcounter{page}{1}
\setcounter{footnote}{0}
\pagestyle{plain}

\vspace{1.0cm}

\section{Introduction}
\label{sec_intro}

Lattice gauge theory is a well-developed discipline within theoretical physics
\cite{MTrev98}.
From its calculations we have learned much about the spectrum of the pure gauge theory, the properties of confinement and deconfinement, as well as topological excitations.  As we explore higher mass and higher spin states, new techniques in calculation will become necessary to evaluate more complicated operators.  This paper explores a technique for evaluating glueball masses from the all-to-all propagator computed from the link variables of a cubic lattice.

Our starting point will be 2+1 dimensional lattice gauge theory.  While simpler than the corresponding 3+1 dimensional theory, there is still plenty of structure in the spectrum to be interesting.  We will focus on the group SU(2) in the quenched approximation, though the discussion will be applicable to other gauge groups.  The primary question is whether it is appropriate to assign the lowest spin compatible with an operator's symmetry group to the state coupling to that operator.  Specifically, we will be interested in the state traditionally labeled with $J^{PC}=0^{-+}$, ie zero spin, negative parity, and positive charge conjugation.  The motivation for this study comes from a robust prediction of the extended Isgur-Paton model, as appearing in 
\cite{RJMTpaper}, \cite{RJthesis},
namely that states with spin 4 and positive charge conjugation degenerate in parity exist at the same mass as the previous spin 0 state.  This prediction extends across all values of the gauge group N.  Considering the {\it{modulo}} 4 ambiguity in spin inherent on a square lattice, determining a method for creating operators of arbitrary spin would now be quite relevant.

In Section~\ref{sec_spinop} we begin examining how to construct an operator of a given spin, first in the continuum and then on the square lattice, highlighting the ambiguity that arises concerning the spin of the state which couples to the operator.  Next, in Section~\ref{sec_triops}, we explore a new technique for constructing operators of arbitrary spin on the lattice.  In Section~\ref{sec_docalc} we present the details of the calculation performed to explore the usefulness of these new operators as well as the numerical results.  Finally, our conclusions are given in Section~\ref{sec_conc}.

\vspace{1.0cm}

\section{The spin of an operator}
\label{sec_spinop}

\subsection{In the continuum}
\label{subsec_cont}

How do we construct an operator of a given spin?  First, we start with some base operator representing the wavefunction of the state in which we are interested, which we may call $\psi$.  In order to be gauge invariant, our operator $\psi$ needs to be a closed loop.  As we are in 2+1 dimensions, we may view our operator as a loop in the plane of, say, this piece of paper.  Pick some axes and denote by $\psi_0$ the loop aligned to the axes.  Then denote by $\psi_\theta$ the loop as rotated by the positive angle $\theta$.  Then, to construct an operator $\Psi_J$ of arbitrary spin J, take the weighted average of all loops $\psi_\theta$, where the weight is the complex phase $e^{iJ\theta}$:
\beq
\Psi_J = \oint d\theta \; e^{iJ\theta} \psi_\theta
\eeq
The symmetry properties of the base operator $\psi_0$ must be taken into account.  If one uses a circularly symmetric loop, then by invariance the only spin which couples is spin 0.  In order to get higher values of spin, loops which are not circularly symmetric must be used.

\subsection{On the lattice}
\label{subsec_onlatt}

\begin{figure}[p]
\centering
\includegraphics[scale=.26]{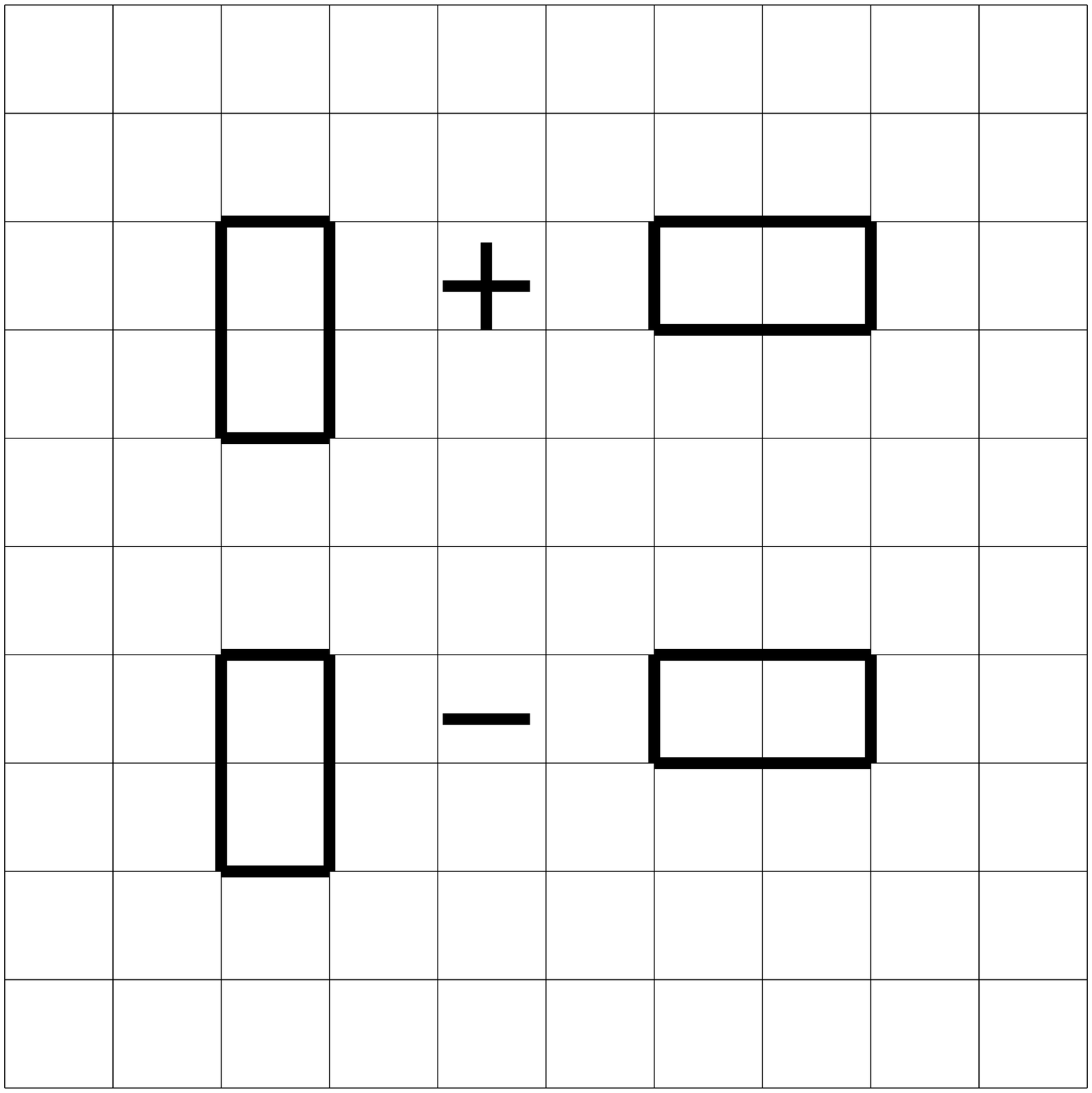}
\caption{The two possible linear combinations of a 1x2 plaquette loop.}
\label{fig_twoplaqs}
\end{figure}

How does the situation change when we go over the lattice?  Instead of having a continuum of spacetime with which to work, we are restricted to constructing operators based on a cubic lattice of points.  Not only does this restrict the amount and variety of loops which we may draw, but also it restricts the available angles at which we may place the loops.  When we go to construct our operator of spin J, we can sum only over the four angles $\theta_j \in \{0,\pi/2,\pi,3\pi/2\}$.
\beq
\Psi_J \rightarrow \sum_j e^{iJ\theta_j} \psi_{\theta_j}
\eeq
The shape of our loop is particularly important for determining the quantum numbers of the state to which it couples.
Consider the simplest non-circularly symmetrical loop available on the lattice, two joined plaquettes (see Figure~\ref{fig_twoplaqs}).  This loop has two distinct angles of orientation, along either the x-axis or the y-axis.  We can get a spin 0 operator by taking the positive linear combination, as shown in the top half of Figure~\ref{fig_twoplaqs}.  If we take the negative combination instead, we get an operator that couples to spin 2.  These operators have positive parity and, as we are restricting ourselves to SU(2), positive charge conjugation.  Note that for these four angles, the phases of a spin 0 and a spin 4 operator are all unity, hence an operator we think couples to a state with spin 0 might really be coupling to a state with spin 4.

\begin{figure}[p]
\centering
\includegraphics[scale=.25]{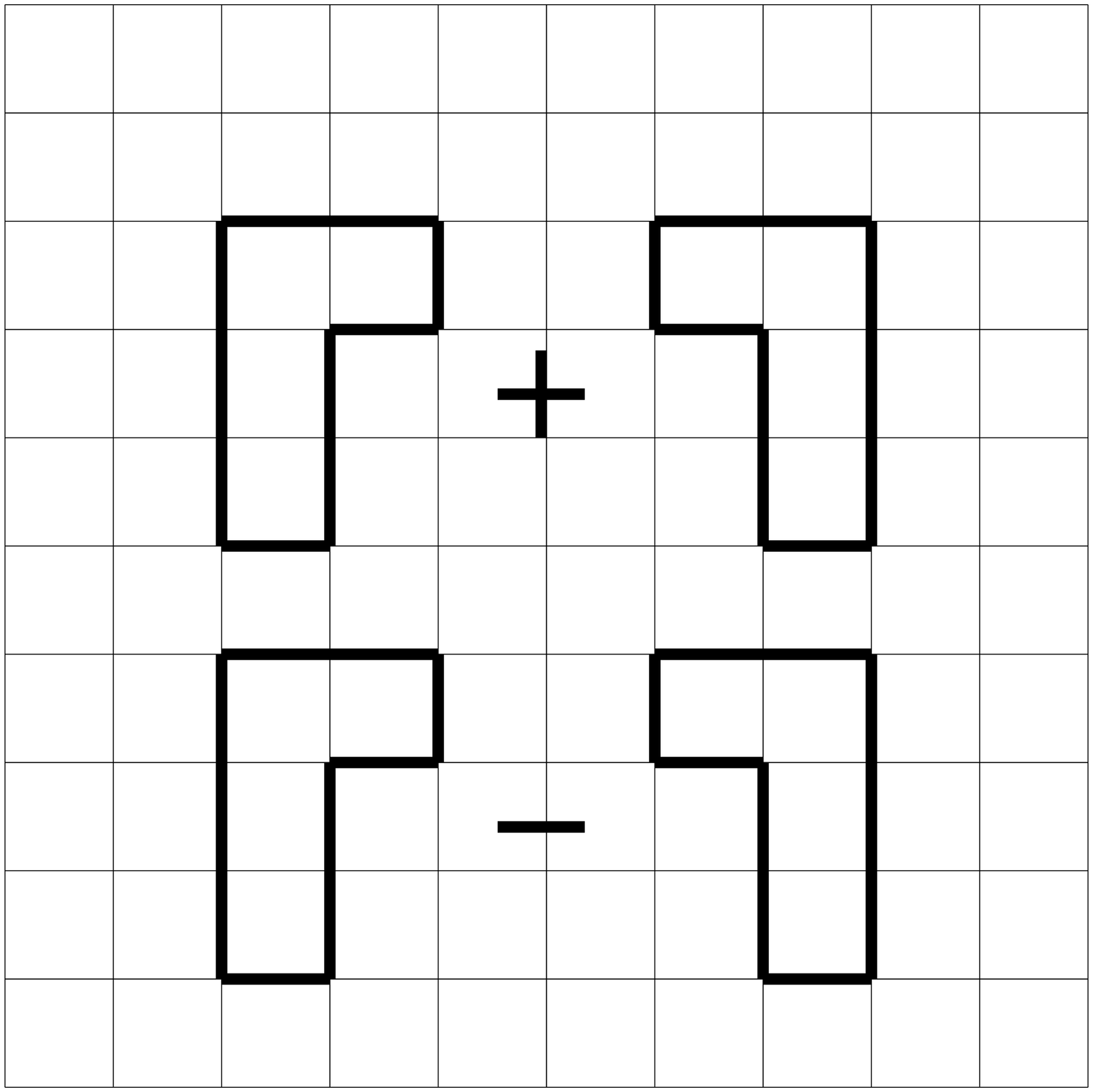}
\caption{The two possible linear combinations of a 3x2 plaquette loop.}
\label{fig_Lplaqs}
\end{figure}

Constructing operators of higher spin and more complicated quantum numbers is an art in itself
\cite{MTpaper98},
 requiring the meticulous application of group theory for the more exotic combinations
\cite{LiuWuChen}.
In our case, we are interested in an operator with spin 0 and negative parity in 2+1 dimensions.  This operator is constructed by first considering a pair of nonsymmetric loops related by a parity transformation, as in Figure~\ref{fig_Lplaqs}.  As no rotation can transform the left hand shape into the right hand shape, the linear combination of this pair can produce an operator of either parity, as given by the addition or subtraction of the loops.  A fully fledged spin 0 operator results from taking the four angular orientations in combination with equal weighting.  In practice, bare lattice links are replaced with ``smeared'' links
\cite{MTsmear}
which follow the axes of the lattice to produce physically significant operators as the lattice spacing goes to zero.

\section{Triangular operators on a square lattice}
\label{sec_triops}

\begin{figure}[p]
\centering
\includegraphics[scale=.8]{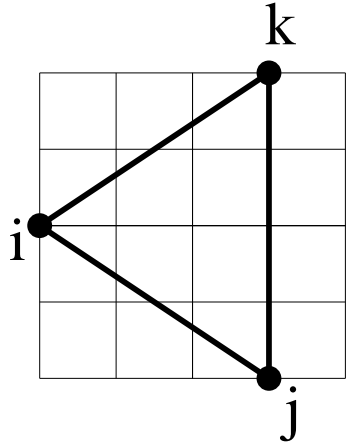}
\caption{The triangle operator.}
\label{fig_triops}
\end{figure}

\begin{figure}[p]
\centering
\includegraphics[scale=.6]{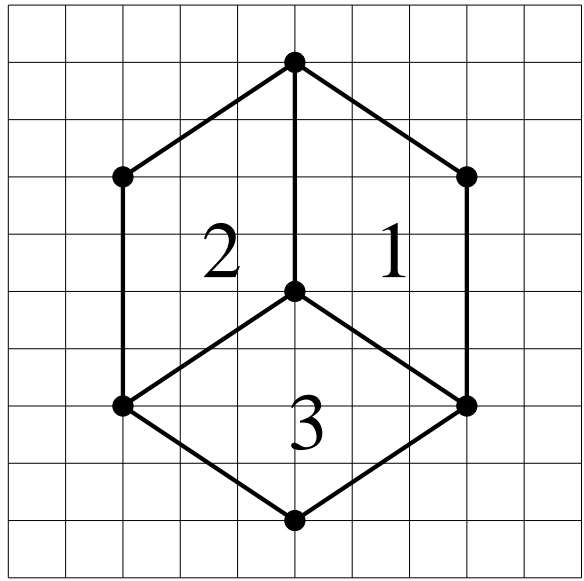}
\caption{The rhomboid operators.}
\label{fig_myops}
\end{figure}

The fundamental difficulty with these operators is that they may only be placed onto the lattice at angles of $\pi/2$.  This property results from using loops based on the plaquette operator.  What we need is a collection of closed loops which may be placed at arbitrary angles on the lattice, or at least at angles other than $\pi/2$.  Perhaps the next simplest shape on the lattice after the square plaquette is the triangle, as in Figure~\ref{fig_triops}.  These triangles may come in various sizes, which we label by half the integral number of links comprising the edge aligned with the lattice axes, eg the triangle in Figure~\ref{fig_triops} is a ``size 2'' triangle.
Do not worry that these triangles are not drawn using the links of the lattice.  Later, in Section~\ref{sec_arblinks} we will show how to compute an effective link between arbitrary lattice points.  Further details on effective links appear in 
\cite{RJthesis}.
These particular triangles are not very useful, coupling effectively only to $J=0$, but from these basic triangles we may build the rhomboid operators of Figure~\ref{fig_myops}, shown also at ``size 2''.  Note that we may place a rhombus at any of three basic orientations on the square lattice, and we may also take the $\pi/2$ analogues, giving a total of six orientations which a rhombus may have.  The vacuum expectation value of the operators labeled 1 and 2 in Figure~\ref{fig_myops} should be equivalent, but the vev for rhombus 3 will be slightly different, as it is not strictly equivalent to a rotation of one of the other orientations.  In practice, following
\cite{RJMTproc},
we normalize all the loops relative to the first one so that they have equivalent vevs, ie
\beq
<\CO_{\theta_1}^2> = c_{\theta_j}^2 <\CO_{\theta_j}^2> \; \mathrm{for} \; j \neq 1 
\eeq
While such a prescription has no formal justification, in practice we find the application of these constants necessary to balance the contributions from the inequivalent orientations.  
From these rhomboid loops we may construct our physical operators by multiplying each loop by the appropriate phase factor $e^{iJ\theta}$ for the spin J.  Notice that now we have operators at the angles
\beq
\theta \in \{ \pi/6,5\pi/6,3\pi/2,\pi/2,7\pi/6,11\pi/6 \}
\eeq 
and so operators of spin 0 and 4 will have different phases for each orientation.

\section{Arbitrary links on the lattice}
\label{sec_arblinks}

In order to construct the triangles and rhombi of the previous section we need to define a ``smeared'' link between {\it{any}} two points of the lattice.  Such an arbitrary link should contain a weighted sum of all paths connecting the two points.  Thus we need to compute a matrix of Greens functions, not just from one lattice site to all others as in fermion calculations, but from {\it{all}} lattice sites to {\it{all other}} lattice sites.  Then the product of appropriate entries in our matrix propagator will evalutate the corresponding operator.

For each timeslice of a given configuration we start by labelling the sites from 1 to $L^2$.  We then construct the $L^2 \times L^2$ matrix $\cM$ whose entries are the SU(2) values of the links of the lattice such that
\beq
\cM_{ij} = \left\{ \begin{array}{ll} 0 & {\textrm{if sites i and j are not nearest neighbors}} \\ U_{ij} & {\textrm{if sites i and j are nearest neighbors}} 
\end{array} \right.
\eeq
where $\cM$ is sparse and Hermitian, since $U_{ij}=U_{ji}^\dagger$.  When $\cM$ multiplies itself, the nonzero entries in the result are the sum of the products of the link variables connecting the sites with the corresponding labels.  In other words, $\cM^2$ is the matrix of all length-2 paths from the timeslice, $\cM^3$ is the matrix of all length-3 paths, {\it{etc}}.  We then define
\beq
K = \frac{1}{1 - \alpha \cM}
\label{eqn_K}
\eeq
and expand as
\beq
K = 1 + \alpha\cM + \alpha^2\cM^2 + ...
\label{eqn_Kseries}
\eeq
for some small parameter $\alpha$ which behaves much like a ``hopping
parameter'' for the gauge field.  Thus, if we can compute K we will have
a matrix whose entries are a weighted sum of all paths connecting any
two sites in the timeslice.  While we might be tempted to compute an
approximate inverse using Equation~\ref{eqn_Kseries} for some finite
series of terms, in practice these operators did not work very well,
even including path lengths up to half the lattice size, and the full inverse from Equation~\ref{eqn_K} was calculated.
Note that there will be a critical value for the parameter, $\alpha_c$, which is related to the reciprocal of the largest eigenvalue of $\cM$.  (See
\cite{RJthesis}
for more information.)  Returning to our triangular operator in Figure~\ref{fig_triops}, we can write the expression for its closed loop as
\beq
\CO_\triangle = K_{ij}K_{jk}K_{ki}
\eeq
and similarly for any closed loop shape which we place on the lattice.  The parameter $\alpha$ essentially controls the amount of ``smearing'' applied to the arbitrary links and will provide an index on our collection of operators during the calculation.

\section{Performing the calculation}
\label{sec_docalc}

As mentioned earlier, we restrict ourselves to SU(2) for this calculation, though the technique will work for any gauge group.  We work with a D=2+1 Euclidean lattice with L=16 at $\beta=6$ in the quenched approximation.  We update the gauge field with the Kennedy-Pendleton heatbath algorithm 
\cite{KPbath}
 supplemented with overrelaxation sweeps 
\cite{ORsweep}
 and occasional global gauge transformations.  Measurements are taken on a set of 1000 configurations separated by 40 heatbath sweeps each.  These are rather low statistics, and so we must view the results cautiously in terms of agreement with the accepted values of the computed spectrum.  Nevertheless, we should have plenty of measurements to evaluate the usefulness of rhomboid operators in distinguishing spins {\textit{modulo}} 4.

The most time consuming part of the calculation is the repeated inversion necessary to solve Equation~\ref{eqn_K} for each $\alpha$ on each timeslice for each configuration measured.  A dedicated algorithm is explored in 
\cite{RJthesis},
 but for this calculation a modified version of the SuperLU program 
\cite{Superlu}
for {\textsc{Matlab}} is used.\footnote{While the analysis happens in {\textsc{Matlab}}, the inversion of the matrix is done in C.}  Once we have K, we can begin evaluating our operators.  We use rhomboid operators at sizes 3,4,5, and 6; size 2 was also computed but never showed much promise.  For each site in a timeslice we compute the values for all six rhomboids at all four sizes and several values of $\alpha$, taking the zero-momentum sum along the way.  Next we compute the normalizing constants $c_\theta$--a measure of how similar the loops actually are is given by how close $c_\theta$ is to unity.  These constants are displayed in Table~\ref{table_consts}.  Note that the four equivalent orientations have constants all near unity, while the two loops of the other orientation require a significant normalization.
While the magnitude of some of the normalization constants might seem alarming, especially at low $\alpha$ and large operator size, these are not the operators which turn out to be useful in extracting effective masses.
As the smearing $\alpha$ increases, we see the constants approach unity for all our operators.

We have the choice of applying our constants $c_\theta$ either to all the loops at hand or just those two loops which are obviously different than the other four, namely loop 3 in Figure~\ref{fig_myops} and its $\pi/2$ analogue.  Applying our constants to the two loops only implies that we expect the $\pi/2$ analogues to affect cancellations (over the duration of the calculation) when computing the J=2 mass so that there is no overlap with the J=0.  Applying our constants to all the loops means we expect no fortuitous cancellations and instead rely on the actual phases applied to our loops to affect the cancellations.  In practice, there is not much difference between the two options, and we shall use the more general application of constants in this calculation.

\begin{figure}[p]
\centering
\includegraphics[scale=.4,trim= 100 200 100 120]{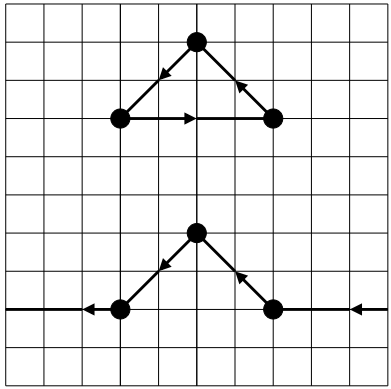}
\caption{Torelon contribution.}
\label{fig_torel}
\end{figure}

\begin{figure}[p]
\centering
\includegraphics[scale=.4]{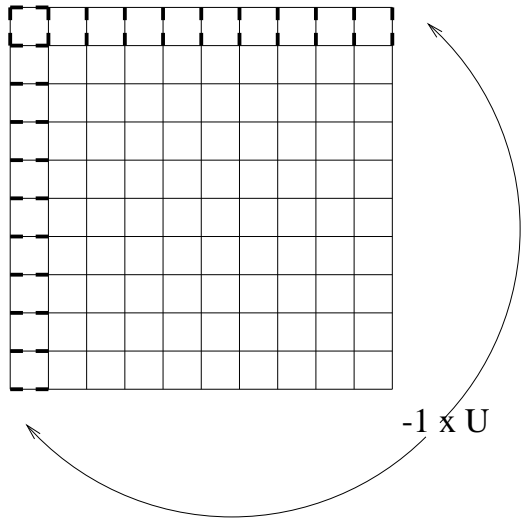}
\caption{Torelon cancellation.}
\label{fig_torcancel}
\end{figure}

Because our expression for the propagator between two lattice sites contains contributions from all paths connecting those two sites, our operators are going to be contaminated with torelon contributions, especially at larger values of $\alpha$.  See Figure~\ref{fig_torel}.  In order to remove these torelon effects, we use the following technique
\cite{MTtorel}.
Noting that paths which wind completely around the timeslice pick up a negative phase contribution when all the links in the x direction at one y coordinate (and {\it{vice versa}}) are multiplied by -1
\cite{MornPear},
we may cancel the torelons by averaging together operators computed with and without the application of the link transformation.  To cancel torelons in all directions, we must compute 4 sets of operators: no link transformation, x-axis link transformation, y-axis link transformation, and both x-and-y-axis link transformation, as in Figure~\ref{fig_torcancel}.  By taking the average of these four operators, only paths which do not wind around the lattice (and so pick up no minus sign) contribute.  Further details may be found in
\cite{RJthesis}.

We now have a set of operators $\CO$ indexed by size, orientation, and $\alpha$.  Operators coupling to spins 0, 2, and 4 are computed by taking the sum of the six loops weighted by the appropriate phase factors.  We can then compute our vacuum-subtracted correlation functions:
\beq
C_J(t) = <\CO_J^\star(t) | \CO_J(0)> - <\CO_J^\star><\CO_J>
\eeq
which will be normalized to $C_J(0)=1$.  Similarly, we can compute the normalized cross-correlation for states of different J:
\beq
C_{{\tilde{J}}J}(t) = {<\tilde{\CO}^\star_{\tilde{J}}(t) | \tilde{\CO}_J>\over \sqrt{<\tilde{\CO}^\star_{\tilde{J}}(t) | \tilde{\CO}_{\tilde{J}}(0)><\tilde{\CO}^\star_J(t) | \tilde{\CO}_J(0)>}}
\label{eqn_overlap}
\eeq
where $\tilde{O}$ implies using the vacuum-subtracted operator.  Taking $t=0$ in Equation~\ref{eqn_overlap} gives us the overlap between the operators, which will be our measure of how well we are distinguishing the J=4 from the J=0.  From these correlation functions we extract the effective masses in units of the lattice spacing:
\beq
m^J_{\mathrm{eff}}(t) = \ln \frac{C_J(t)}{C_J(t+1)}
\eeq
Statistical errors are calculated using a standard jackknife analysis
\cite{jack} with 20 bins.

\begin{figure}[!h]
\centering
\includegraphics[width=\textwidth]{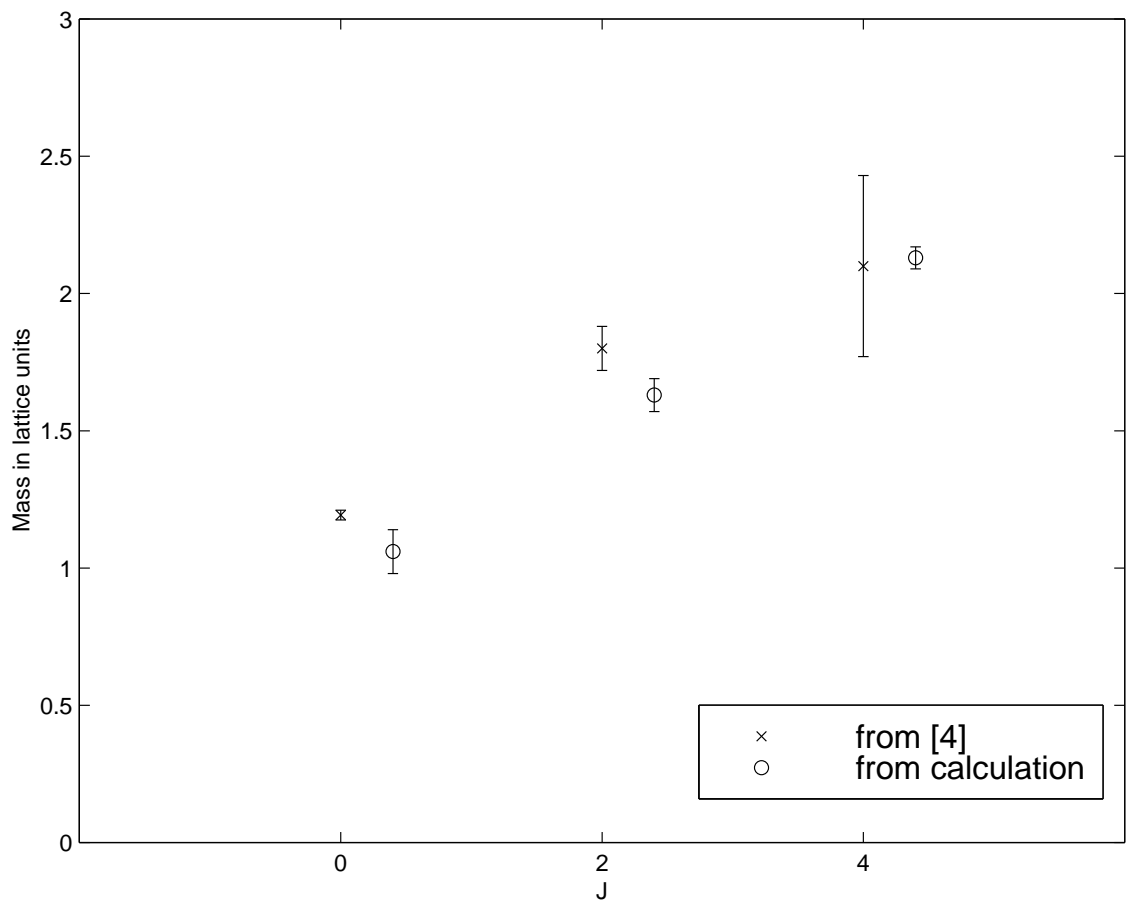}
\caption{The results of the present calculation compared to the masses from~\cite{MTpaper98}.}
\label{fig_latres}
\end{figure}

Working with a coarse lattice, we expect the effective mass of the lightest state, the $0^+$, to be dominated by excitations at one lattice spacing, and so we look for its mass to appear at a lattice spacing of two or three.  For the heavier states, the $2^+$ and the $4^+$, the correlation functions have already dropped down to the level of the statistical noise by two lattice spacings, and so we must select our answer from the effective masses at one lattice spacing.  The effective masses are presented in Tables~\ref{table_mass0p4} through \ref{table_mass4p4}.  For J=2, we notice that all the overlaps are less than 2\%, and so we select the lightest mass as our result.  For J=4, we are more concerned about the magnitude of the overlap with the J=0 than we are with selecting the lightest mass, and so we take the mass of the state with the smallest overlap as our result.  The chosen masses are in boldface in the tables and are presented graphically alongside the results from \cite{MTpaper98} in Figure~\ref{fig_latres}, where the $0^-$ of \cite{MTpaper98} is compared with the $4^+$ from the present calculation.

From the figure we note several interesting observations.  First, while the size of the errorbars from~\cite{MTpaper98} increase with the magnitude of the mass, the current statistical errorbars actually decrease.
Secondly, we remark on the agreement in the J=0 and J=2 channels; while slightly lower than the accepted values for these states, the computed masses compare quite well when we remember that we are using an entirely different technique than previous studies, and perhaps were the current technique extended across many values of $\beta$ we would see agreement in the continuum limit, but that is pure speculation.  Finally, we note the ``agreement'' between the $0^-$ and the $4^+$.  How can we call this agreement?  Because in D=2+1 we have the curious property of parity doubling, which holds that any state with nonzero spin in 2+1 dimensions exists as a degenerate parity doublet.  Thus, when we calculate the mass of the $4^+$ we are also determining the mass for the $4^-$.  And so, we claim that the state traditionally labelled $0^-$ in lattice calculations 
\cite{MTpaper98} actually has a spin of 4.

\section{Conclusions}
\label{sec_conc}

There is much progress left to be made in resolving the spin structure of the spectrum for even the simple case of SU(2) in 2+1 dimensions, let alone the more relevant case of SU(3) in 3+1 dimensions.  Nonetheless, by using the technique of arbitrary lattice links we can start considering operators beyond those constrained by construction from elementary plaquettes.  Any closed shape which may be drawn using the points of the lattice becomes a potential glueball operator.  Shapes with more varied symmetry properties than those currently used are now available.  And with these new shapes we can create operators that are not affected by the {\textit{modulo}} 4 spin ambiguity of current operators.

To calculate these new operators requires the repeated inversion of a moderately large matrix.  This calculation was painfully slow in its execution, perhaps by being written in an interpreted language rather than a compiled one.  No attempt was made to optimize the code beyond picking the fastest readily available inversion routine.  Substantial improvements may be made by implementing the algorithm found in 
\cite{RJthesis}
and using an optimizing compiler, as well as incorporating information from previous inversions when calculating the next.  Improvements may also be made in understanding the systematic errors of this technique, of which we know little.

Even with such work remaining to be done, the usefulness of these rhomboid operators seems apparent.  The ability to distinguish spins previously obscured by the $\pi/2$ angular constraint of the square lattice will become more important as the heavier reaches of the spectrum are explored.  These spin resolutions are important in evaluating how well a more phenomenological model explains the lattice spectrum, as in 
\cite{RJMTpaper}.
With further development, arbitrary link operators might come to rival traditional smeared operators in terms of speed and accuracy, at least when distinguishing 0 from 4.

\section*{Acknowledgments}
This research was performed as part of the author's doctoral research.  The author is grateful to his supervisor, Mike Teper, for suggesting the matrix method described in Section~\ref{sec_arblinks} and the technique for torelon removal, as well as for many useful discussions.
The author thanks the Rhodes Trust and the Astrophysics Department of Oxford University for financial support.

\newpage

\begin{table}[p]
\centering
\begin{tabular}{|c|cccc|} \hline

loop & & \multicolumn{2}{c}{size}  & \\
\# & 3 & 4 & 5 & 6 \\\hline

 &  &  \multicolumn{2}{c}{$\alpha=.1$}    & \\\hline
1 & 1.0000e+00 & 1.0000e+00 & 1.0000e+00 & 1.0000e+00 \\
2 & 1.0082e+00 & 9.6166e-01 & 1.0237e+00 & 1.0161e+00 \\
3 & 1.8708e+01 & 1.3566e+02 & 1.3982e+05 & 1.4130e+07 \\
4 & 1.0055e+00 & 9.5868e-01 & 1.0346e+00 & 1.0211e+00 \\
5 & 9.9742e-01 & 9.7717e-01 & 1.0386e+00 & 1.0196e+00 \\
6 & 1.8997e+01 & 1.4096e+02 & 1.4026e+05 & 1.4339e+07 \\\hline

 &  &  \multicolumn{2}{c}{$\alpha=.15$}    & \\\hline
1 & 1.0000e+00 & 1.0000e+00 & 1.0000e+00 & 1.0000e+00 \\
2 & 9.9819e-01 & 9.5151e-01 & 1.0282e+00 & 1.0201e+00 \\
3 & 4.6171e+00 & 1.7857e+01 & 2.5579e+03 & 8.5268e+04 \\
4 & 1.0044e+00 & 9.5565e-01 & 1.0428e+00 & 1.0210e+00 \\
5 & 9.9262e-01 & 9.7976e-01 & 1.0322e+00 & 1.0253e+00 \\
6 & 4.7088e+00 & 1.8702e+01 & 2.5747e+03 & 8.6693e+04 \\\hline

 &  &  \multicolumn{2}{c}{$\alpha=.2$}    & \\\hline
1 & 1.0000e+00 & 1.0000e+00 & 1.0000e+00 & 1.0000e+00 \\
2 & 9.8341e-01 & 1.0141e+00 & 1.0188e+00 & 1.0209e+00 \\
3 & 1.9445e+00 & 4.8168e+00 & 1.2366e+02 & 1.5746e+03 \\
4 & 9.9933e-01 & 9.9477e-01 & 1.0340e+00 & 1.0198e+00 \\
5 & 9.8311e-01 & 1.0596e+00 & 1.0502e+00 & 1.0290e+00 \\
6 & 1.9989e+00 & 4.9811e+00 & 1.2663e+02 & 1.6053e+03 \\\hline

 &  &  \multicolumn{2}{c}{$\alpha=.25$}    & \\\hline
1 & 1.0000e+00 & 1.0000e+00 & 1.0000e+00 & 1.0000e+00 \\
2 & 9.7862e-01 & 9.6075e-01 & 9.9969e-01 & 9.7553e-01 \\
3 & 1.0331e+00 & 1.0580e+00 & 3.3177e+00 & 1.5299e+01 \\
4 & 9.9450e-01 & 9.7818e-01 & 9.7776e-01 & 9.8693e-01 \\
5 & 9.5446e-01 & 9.7849e-01 & 1.0289e+00 & 1.0236e+00 \\
6 & 1.1033e+00 & 1.1611e+00 & 3.3920e+00 & 1.5382e+01 \\\hline

\end{tabular}
\caption{Loop normalization constants.}
\label{table_consts}
\end{table}

\begin{table}[p]
\centering
\begin{tabular}{|c|cccc|} \hline
t =  &  & 1 & 2 & 3 \\\hline

size &  &  \multicolumn{2}{c}{$\alpha=.1$}    & \\\hline
3 & &  1.37(03) & 1.32(10) & 1.58(50)  \\
4 & &  1.51(04) & 1.50(18) & 1.76(84)  \\
5 & &  1.49(03) & 1.41(12) & -         \\
6 & &  1.37(03) & 1.27(10) & 2.06(87)  \\\hline

 &  &  \multicolumn{2}{c}{$\alpha=.15$}    & \\\hline
3 & &  1.35(03) & 1.30(10) & 1.55(45) \\
4 & &  1.45(03) & 1.39(14) & 1.64(66) \\
5 & &  1.44(03) & 1.36(10) & -        \\
6 & &  1.35(03) & 1.26(10) & 1.97(75) \\\hline

 &  &  \multicolumn{2}{c}{$\alpha=.2$}    & \\\hline
3 & &  1.34(03) & 1.28(10) & 1.52(40) \\
4 & &  1.38(03) & 1.27(10) & 1.57(52) \\
5 & &  1.37(03) & 1.28(08) & -        \\
6 & &  1.33(03) & 1.23(09) & 1.84(59) \\\hline

 &  &  \multicolumn{2}{c}{$\alpha=.25$}    & \\\hline
3 & &  1.49(04) & 1.20(07) &  1.21(32)  \\
4 & &  1.42(04) & {\bf{1.06(08)}} &  1.04(24)  \\
5 & &  1.46(04) & 1.17(08) &  1.40(41)  \\
6 & &  1.49(04) & 1.23(08) &  1.77(65)  \\\hline

\end{tabular}
\caption{Masses for $0^{+}$ after torelon removal.}
\label{table_mass0p4}
\end{table}

\begin{table}[p]
\centering
\begin{tabular}{|c|ccc|c|} \hline
t =  & & 1 & 2 & $|<\CO_2 | \CO_0>|$ \\\hline

size  & &  \multicolumn{2}{c}{$\alpha=.1$} & \\\hline
3 & &  2.52(08) & 1.93(05) & .005(4) \\
4 & &  2.37(07) & 1.62(03) & .007(3) \\
5 & &  2.52(09) & 2.39(07) & .008(6) \\
6 & &  2.32(06) & -        & .006(7) \\\hline

  & &  \multicolumn{2}{c}{$\alpha=.15$} & \\\hline
3 & &  2.47(08) & 2.10(50) & .005(4) \\
4 & &  2.06(06) & 1.43(19) & .011(3) \\
5 & &  2.32(08) & 1.99(42) & .011(5) \\
6 & &  2.29(06) & -        & .007(7) \\\hline

  & &  \multicolumn{2}{c}{$\alpha=.2$}  & \\\hline
3 & &  2.44(08) & 2.85(1.15)& .004(4) \\
4 & &  1.68(05) & 1.36(13)  & .016(4) \\
5 & &  1.97(08) & 1.64(23)  & .014(5) \\
6 & &  2.30(06) & -         & .008(6) \\\hline

  & &  \multicolumn{2}{c}{$\alpha=.25$} &  \\\hline
3 & &  2.08(07) & -         & .009(6) \\
4 & &  {\bf{1.63(06)}} & 3.13(1.11)& .015(7) \\
5 & &  1.79(07) & -         & .015(6) \\
6 & &  2.06(07) & -         & .010(4) \\\hline

\end{tabular}
\caption{Masses for $2^{+}$ after torelon removal.  The last column is the overlap with the $0^{+}$.}
\label{table_mass2p4}
\end{table}

\begin{table}[p]
\centering
\begin{tabular}{|c|ccc|c|} \hline
t = & & 1 & 2 & $|<\CO_4 | \CO_0>|$ \\\hline

size  & &  \multicolumn{2}{c}{$\alpha=.1$} & \\\hline
3 & &  2.47(08) & -        & .39(1) \\
4 & &  2.15(05) & 1.91(28) & .35(1) \\
5 & &  2.74(09) & 2.01(99) & .27(1) \\
6 & &  2.32(04) & 1.70(53) & .13(1) \\\hline

  & &  \multicolumn{2}{c}{$\alpha=.15$} & \\\hline
3 & &  2.73(10) & -        & .27(1) \\
4 & &  2.16(06) & 1.89(27) & .24(1) \\
5 & &  2.51(07) & 1.54(46) & .17(1) \\
6 & &  2.23(04) & 1.64(47) & .05(1) \\\hline

  & &  \multicolumn{2}{c}{$\alpha=.2$}  & \\\hline
3 & &  2.99(12) & -        & .11(1) \\
4 & &  1.99(05) & 1.63(21) & .11(1) \\
5 & &  2.00(05) & 1.27(18) & .14(1) \\
6 & &  {\bf{2.13(04)}} & 1.58(39) & .007(3) \\\hline

  & &  \multicolumn{2}{c}{$\alpha=.25$} &  \\\hline
3 & &  2.66(08) & -        & .03(1) \\
4 & &  2.11(07) & 2.77(93) & .38(1) \\
5 & &  2.11(05) & 1.93(28) & .05(1) \\
6 & &  2.06(05) & -        & .49(1) \\\hline

\end{tabular}
\caption{Masses for $4^{+}$ after torelon removal.  The last column is the overlap with the $0^{+}$.}
\label{table_mass4p4}
\end{table}

\newpage

\end{document}